\documentclass[10pt,preprint2]{aastex}

\newcommand{\revision}[1]{#1}

\newcommand{\Msun   }{{M}_{\sun}}
\newcommand{\Rsun   }{{R}_{\sun}}
\newcommand{\Fig    }{Figure~}

\newcommand{\Sec    }{Section~}
\newcommand{\eboo   }{{$\eta$~Boo~}}
\newcommand{\eboop  }{{$\eta$~Boo}}
\newcommand{\ebootis}{{$\eta$~Bootis~}}

\slugcomment{Revision 2.}

\shorttitle{\eboo Pulsation Data Explained with Turbulence}
\shortauthors{Straka, Demarque, Guenther, Li \& Robinson}

\begin{document}
  
\title{Space and Ground Based Pulsation Data of \ebootis Explained
       with Stellar Models Including Turbulence}

\author{Christian~W.~Straka\altaffilmark{1},
        Pierre~Demarque    \altaffilmark{1},
	D.~B.~Guenther     \altaffilmark{3},\\
        Linghuai~Li        \altaffilmark{1}, and
        Frank~J.~Robinson  \altaffilmark{2}
        }
\altaffiltext{1}{Department of Astronomy, Yale University, P.O. Box 208101, New Haven, CT 06520-8101;
                 straka@astro.yale.edu, li@astro.yale.edu, demarque@astro.yale.edu}
\altaffiltext{2}{Department of Geology and Geophysics, Yale University,  New Haven, CT 06520-8101;
                 marjf@astro.yale.edu}
\altaffiltext{3}{Institute for Computational Astrophysics, Department of Astronomy and Physics,
                 Saint Mary's University, Halifax, N.S., Canada, B3H 3C3;
                 dguenther@eastlink.ca}

\begin{abstract}
The space telescope MOST is now providing us
with extremely accurate low frequency $p$-mode oscillation data
for the star \eboop. We
demonstrate in this paper that these data, when combined with
ground based measurements of the high frequency $p$-mode spectrum,
can be reproduced with
stellar models that include the effects of turbulence in their
outer layers. Without turbulence, the $l=0$ modes of
our models deviate from either
the ground based or the space data by about $1.5$--$4\,\mu$Hz. This
discrepancy can be completely removed by including
turbulence in the models and we can exactly match $12$ out of $13$
MOST frequencies that we identified as
$l=0$ modes in addition to $13$ out of
$21$ ground based frequencies within their observational $2\sigma$
tolerances. \revision{The better agreement between model
frequencies and observed ones depends for the most part on the turbulent
kinetic energy which was taken from a 3D convection simulation
for the Sun.}
\end{abstract}

\keywords{  stars: evolution --- stars: individual (\eboop) --- stars: oscillations --- turbulence}

\section{Introduction}
\label{sec:intro}
Until now, high precision stellar modeling of the outer
stellar layers was needed solely for helioseismic studies.
In order to reproduce the observed solar $p$-mode oscillation
frequencies, \markcite{2002ApJ...567.1192L}{Li} {et~al.} (2002) demonstrated that
the proper inclusion of turbulence improves the observed
solar high frequency $p$-modes from a maximum deviation
of $15\,\mu$Hz at $4000\,\mu$Hz for a model without turbulence to
$5\,\mu$Hz for a model with turbulence.

The inclusion of turbulence is a twofold problem. It
consists of realistically modeling turbulent convection in the
outer layers and then including simulation data in
stellar models. 
Semianalytical models for turbulent
convection have been proposed by
\markcite{1990A&A...227..282C,1996ApJ...467..385C}{Canuto} (1990, 1996). His main
idea is to include a full turbulence spectrum.
Canuto's model has been
included in stellar codes by
\markcite{1991ApJ...370..295C}{Canuto} \& {Mazzitelli} (1991) and \markcite{1996ApJ...473..550C}{Canuto}, {Goldman}, \&  {Mazzitelli} (1996). The
free parameters in the semi-analytical model were
derived from laboratory experiments of
incompressible convection and extrapolated to stellar conditions.
Using this approach, the superadiabatic peak is much higher than
that of the standard solar model (SSM), while the derived
$p$-modes are closer to the observed values than those from SSMs
\markcite{1993ApJ...402..733P}({Paterno} {et~al.} 1993).

In another approach, three-dimensional (3D)
large eddy simulations of deep compressible convection
have been first performed by \markcite{1989ApJ...336.1022C}{Chan} \& {Sofia} (1989).
While it is not possible to resolve the
full turbulence spectrum in 3D simulations, the
advantage of these simulations is that they are
essentially parameter-free, provided that the employed
subgrid model does not significantly modify the properties of the
large scale dynamics.
In the early studies, no account was taken of
the radiative transfer in the 3D simulation, and therefore
the stellar models that included a
parametrized convective flux from the simulations
showed a larger discrepancy with observed solar
$p$-mode frequencies \markcite{1992ApJ...397..701L,1993PhDT........52L}({Lydon}, {Fox}, \&  {Sofia} 1992; {Lydon} 1993).

Part of these limitations were overcome by
\markcite{1995ApJ...442..422K,1996ApJ...461..499K}{Kim} {et~al.} (1995, 1996) who
employed the diffusion approximation for the radiation
field but consequently could not include the optically
thin part of the SAL.
Later, \markcite{1997scor.proc..131K,1998ApJ...496L.121K}{Kim} \& {Chan} (1997, 1998)
employed the Eddington approximation for the radiation field,
included a realistic equation of state and radiative opacities.
Their simulation spanned 5.5 pressure scale heights and included all of the SAL.
\markcite{1999ApJ...517..510D}{Demarque}, {Guenther}, \&  {Kim} (1999) mimicked the effects of
the simulations in calibrated solar models by increasing
the opacity coefficient $\kappa$ which decreased the
discrepancy between observed and model $p$-mode frequencies.

Using different numerical methods to solve the convective
and radiative equations, \markcite{1998ApJ...499..914S}{Stein} \& {Nordlund} (1998) also performed
full 3D simulations, incorporating LTE radiative
transfer and a realistic equation of state. The simulation
included the entire SAL and spanned a total of 11 pressure scale heights.
\markcite{1999A&A...351..689R}{Rosenthal} {et~al.} (1999) used averages of
\markcite{1998ApJ...499..914S}{Stein} \& {Nordlund} (1998)'s hydrodynamical simulations 
to match the simulation to
an envelope that was constructed with a standard mixing length
envelope code. These patched models showed a better agreement
with the observed $p$-mode frequencies than earlier models.

In this paper we make use of the recently
performed 3D simulations
of fully compressible hydrodynamics
by \markcite{2003MNRAS.340..923R,2004MNRAS.347.1208R}{Robinson} {et~al.} (2003, 2004). These
efforts build on the earlier work by \markcite{1998ApJ...496L.121K}{Kim} \& {Chan} (1998).
While resolving the SAL and covering $7.4$ pressure scale heights in the
vertical domain, they yielded results in agreement with
\markcite{1998ApJ...499..914S}{Stein} \& {Nordlund} (1998).
Robinson et al.'s studies showed that artifacts of the boundary conditions
had affected the \markcite{1998ApJ...496L.121K}{Kim} \& {Chan} (1998)
3D simulations and they ascertained the resolution
and domain sizes needed to
yield physically realistic results. Using averages for
the turbulent pressure and turbulent kinetic energy
taken from these simulations \markcite{2002ApJ...567.1192L}{Li} {et~al.} (2002) included these
effects on the stellar structure of the one dimensional
models within the mixing length theory (MLT) framework. So far,
the best match to observed solar $p$-mode frequencies has
been achieved with these methods.

The space mission MOST\footnote{MOST (Microvariability \&
Oscillations of STars) is a Canadian Space Agency mission,
jointly operated by Dynacon Inc., the University of Toronto
Institute for Aerospace Studies and the University of
British Columbia, with the assistance of the University of Vienna.}
\markcite{2003PASP..115.1023W}({Walker} {et~al.} 2003)
is now providing us with new low frequency $p$-modes for \eboo
\markcite{2005ApJ...........G}({Guenther} {et~al.} 2005).
With the ground based measurements of the high frequency part
from \markcite{2003AJ....126.1483K}{Kjeldsen} {et~al.} (2003),
which are sensitive to the outer stellar layers of the star,
we demonstrate that the combined
data set can be \revision{matched within the errors}
when we include turbulence in the outer layers of this star.
The ground based data by
\markcite{2005A&A...434.1085C}{Carrier}, {Eggenberger}, \&  {Bouchy} (2005) are also discussed.



\section{Stellar Models}
\label{sec:models}
\subsection{Turbulence}
\label{sec:turb}
For the Sun, \markcite{2002ApJ...567.1192L}{Li} {et~al.} (2002) have devised a method to
include the effects of turbulence obtained from 3D
hydrodynamic simulations \markcite{2003MNRAS.340..923R}({Robinson} {et~al.} 2003)
by including both the turbulent pressure and
the turbulent kinetic energy \revision{into the 1D stellar model}
within the framework of standard MLT. This method produces
$p$-mode frequencies that match the measured solar spectrum
better than an SSM without turbulence.

We slightly modify the techniques of Li et al. to enable us 
to apply the 3D turbulence data for the Sun to our
model of \eboop. Until we have completed a full 3D simulation
for the outer layers of \eboop, we make use of
the solar data and appropriately shift the data to apply
it at the correct depth in \eboop.
This shifting is motivated by an expected
characteristic found in all our 3D simulations: namely that
the SAL peak closely coincides with the turbulent
pressure peak.

\begin{figure}[t]
\begin{center}
\includegraphics[height=0.45\textwidth,angle=-90]{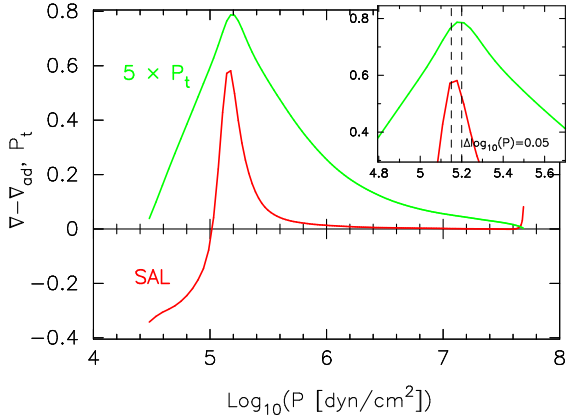}
\end{center}
\caption{Averaged quantities taken from a 3D turbulent
convection simulation for the Sun. The peak of the SAL
coincides with the peak of turbulent pressure. The inset
magnifies the peak locations to show their relative positions.
\label{fig:peak}}
\end{figure}

This is verified for 3D simulations of four different evolutionary stages of the Sun
(ZAMS, current Sun, subgiant and giant). In each case
the peak of the turbulent pressure 
closely coincides with the peak of the SAL.
\Fig\ref{fig:peak}
illustrates this property for the Sun.
The offset between the peaks measured in pressure
difference is always smaller than
$\Delta\!\log_{10}(P\,[\mathrm{dyn/cm^2}]) = 0.1$.
Therefore, in order to apply the solar turbulence data
to \eboo we match the turbulent pressure peak from
the 3D simulation data with the SAL of the one dimensional
model with a small offset of
$\Delta\!\log_{10}(P\,[\mathrm{dyn/cm^2}]) = 0.0758$, a
mean value derived from the four solar models. 
This ensures that the solar turbulence data is
applied at the correct depth of the \eboo model.

\eboo exhibits
a slightly higher effective temperature of $T_\mathrm{eff}=6028\,$K
which amounts to a 18\% higher flux compared to the Sun. The
surface gravity of \eboo is a factor of four lower.
Both differences can change the relative
strength of the turbulent pressure and turbulent kinetic
energy in the 3D simulation but we do not have enough
simulations available yet in order to extract sensible scaling
relations. Therefore, no further scaling has been performed.

The refined treatment of the outer stellar layers
has been implemented in the
Yale Stellar Evolution Code (YREC). The numerical methods
and main physics included are described by
\markcite{1988PhDT........14P}{Pinsonneault} (1988) and \markcite{1992ApJ...387..372G}{Guenther} {et~al.} (1992).
The most recent improvements other than the
inclusion of turbulence in the outer layers
include new updates to the equation of state
\markcite{rogers2001}(OPAL 2001 EOS, {Rogers} 2001).

The high spatial resolution needed for the solar models is also
required for the \eboo models. A model typically consists of
about $4500$ grid points which are distributed in order to
give smooth runs of all variables. In contrast to the solar
models we find that a more stringent time-stepping is needed
for \eboo in order to yield resolution independent
$p$-mode frequencies. This is
most likely due to {\eboop}'s more advanced evolutionary
stage. This demands us to advance the model through
at least $2500$ time steps.

\subsection{Starting Model}
\label{sec:start}
Our model construction starts from a model for \eboo that
has been selected with the \emph{quantified dense grid method}
(QDG) developed by \markcite{2004ApJ...600..419G}{Guenther} \& {Brown} (2004). The
search performed for \eboo is described in detail by
\markcite{2005ApJ...........G}{Guenther} {et~al.} (2005). Their best fit model
is selected from an extended search with different input
parameters for hydrogen $X = (0.69,0.71)$,
metallicity $Z = (0.02,0.03,0.04)$
and stellar masses between $1.4\,\Msun$ and $1.9\,\Msun$
with a fine grid resolution of $0.005\,\Msun$.
Along each evolution track of the models, $p$-mode frequency spectra for
the $l=0,1,2,3$ modes have been calculated from radial order $n=1$ to
the acoustic cutoff frequency. 

A total of $3\times 10^7$ model frequency spectra have been
compared with eight selected MOST frequency measurements that
were judged to be the most likely members of the $l=0$
$p$-mode sequence.
The agreement between model spectra and observation is ascertained
with the $\chi^2$ formulation \markcite{2004ApJ...600..419G}({Guenther} \& {Brown} 2004). The best
model consists of a mass of $1.71 \pm 0.05\,\Msun$, $(X,Z) = (0.71,0.04)$,
a mixing length of $1.8$ and no element
diffusion\footnote{See \markcite{2004ApJ...612..454G}{Guenther} (2004) for an discussion
about diffusion in \eboop.} at an evolution age of $2.40 \pm 0.03 \,$Gyrs.
This best model with a $\chi^2 < 1.4$
was constrained only by the 8
MOST $p$-modes. No other constraints, such as composition, surface
temperature, or luminosity were used. Regardless, their model which best
fits the oscillation data also lies within $1\sigma$ of the
observationally derived effective temperature, luminosity, and metal
abundance.

\revision{A new interferometric measurement of {\eboop}'s radius
is now available \markcite{2005A&A...436..253T}({Th{\' e}venin} {et~al.} 2005), yielding
a radius of $R/{\Rsun} = 2.68 \pm 0.05$. The best model
selected with the QDG search technique is fully consistent with this value,
since it possesses a radius of $R/\Rsun = 2.6842$.
\markcite{2005ApJ...........G}{Guenther} {et~al.} (2005) show that there is no other model within
the searched parameter-space that fits both the MOST data and, in addition,
the effective temperature and luminosity as derived 
observationally by \markcite{2003A&A...404..341D}{Di Mauro} {et~al.} (2003). Thus, the
new interferometric radius measurement does not give any
additional constraint to our modeling, nevertheless it is an
essential requirement that our models are consistent with this
observationally determined radius.}

\begin{figure}[tp]
\begin{center}
\includegraphics[width=0.46\textwidth,angle=0]{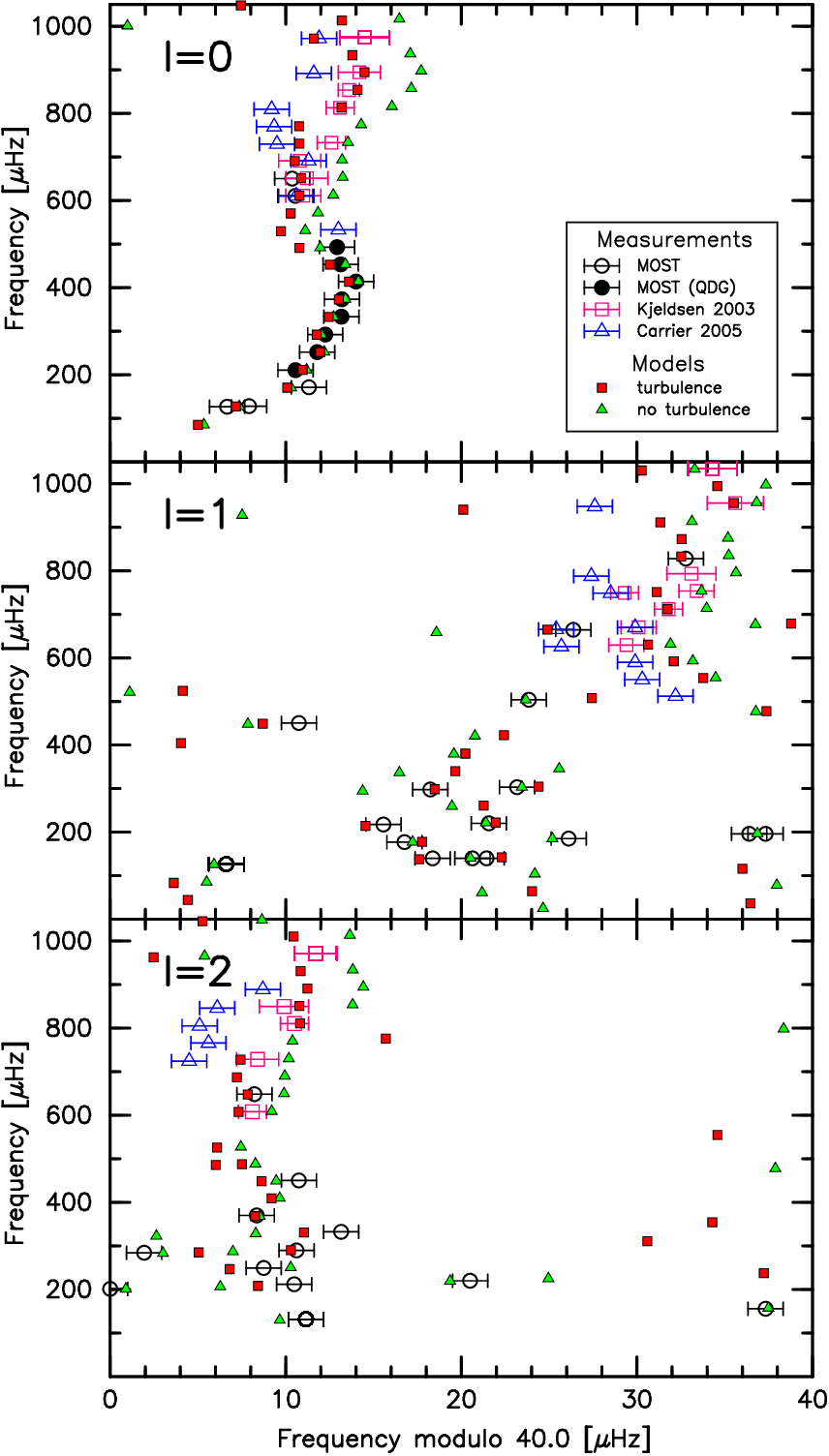}
\end{center}
\caption{Echelle diagram showing the non-adiabatic
$p$-mode frequencies derived from a best-fit theoretical 
model without turbulence (triangles) in comparison to a
model with turbulence (squares) on top of the ground
and space based observational measurements. The eight
selected MOST measurements used in the QDG search are marked with
filled circles.
The quoted error bars correspond to a $2\,\sigma$ deviation.
\label{fig:ech1}}
\end{figure}

The pulsation spectra
computed for this paper are calculated from a model with
exactly the same input parameters but using a slightly
different version of YREC that incorporates the newer
OPAL 2001 EOS. For an evolutionary age of $2.409\,$Gyrs
we achieve a favorable fit of $\chi^2 < 1.0$.
This model is shown in an echelle diagram in \Fig\ref{fig:ech1}
(green triangles). All $\chi^2$-numbers are
calculated with an adopted model uncertainty of $0.05\,\mu$Hz,
and with the exact $1\sigma$ uncertainty as quoted by
the authors for their individual measurements.

It is important to note that none of the models
without turbulence -- searched for
giving a good $\chi^2$-fit for eight selected
MOST low $p$-mode frequencies --
match the $l=0$ ground based frequency
measurements satisfactorily.
The model frequencies appear to be at slightly
higher folded frequencies with a difference 
that increases from $1.5\,\mu$Hz at $600\,\mu$Hz
to $4\,\mu$Hz at $900\,\mu$Hz
(\Fig\ref{fig:ech1}). These differences between
model and observations are marginally within observational uncertainty
up to $700\,\mu$Hz, but above $700\,\mu$Hz they are
significant.

\subsection{Model Calibration}
\label{sec:cal}
Next we construct
a neighboring model to our previously
defined best fit model that yields the same good agreement to
the observed MOST $p$-mode frequencies and
in addition includes the effects of turbulence in the
outer layers. There is a fundamental difference in
calibrating models to \eboo compared to calibrating models
to the Sun. For the latter
we know the age to high precision. Therefore,
a solar model is calibrated by evolving the model to the
exact same age and changing two unknown stellar parameters,
i.e., mixing length and hydrogen mass fraction. 
To first order, the luminosity depends on the hydrogen mass fraction
and the effective temperature is most sensitive to the
mixing length parameter.
By attempting to follow a similar procedure
with \eboo we are faced with the difficulty that neither the age
nor the mass of this star are known.
In order to fit a specific locus in the
HRD we can, e.g., hold the mixing length constant and
only vary the hydrogen content and the age. However, this choice
is arbitrary and we could with equal justification have held either
the hydrogen content or the age constant, while varying the remaining
other two.

To find a proper calibration method we look
at the effect
on the $p$-mode frequencies of
changing \emph{one} of the three
free parameters (mixing length, hydrogen content or age)
while keeping the other two fixed.
Changing the age simultaneously
alters \emph{all} frequencies,
thus altering the age would destroy our good match with the
MOST data. Turning this finding around, we can view
the QDG search for a match of the lower frequency $p$-modes
as a method \revision{for} finding the age and locus in the HRD of
\eboop. This finding is supported by the more
rigorous analysis made in \markcite{2005ApJ...........G}{Guenther} {et~al.} (2005) where
it is shown that the low frequency $p$-modes anchor
the interior structure, hence mass and age, effectively.

With the age and mass being fixed by
the low frequency $p$-modes, we conclude that
the calibration of a model with turbulence has to be performed
the same way as for the Sun by changing the mixing length
parameter and the hydrogen mass fraction.
However, the age parameter \revision{could} be used to fine-tune and
improve the $\chi^2$-fit of the combined low and
high- frequency data sets, 
a possibility not taken advantage of here.

\section{Results}
\label{sec:results}
We now put together the different elements discussed in the
previous sections in order to derive the pulsation spectrum
of a model for \eboo that includes turbulence. The mass, metallicity,
age, luminosity and effective temperature of \eboo is derived with
the QDG search technique outlined in \Sec\ref{sec:start}
yielding a model in the subgiant evolutionary phase at
age $2.409\,$Gyrs, mass of $1.710\,\Msun$, metallicity of $Z=0.04$
and mixing length of $1.8$ that fits eight selected
$p$-mode observations of MOST with $\chi^2 < 1.0$.
The non-adiabatic $p$-mode
frequencies for this model are calculated with
Guenther's pulsation code JIG \markcite{1994ApJ...422..400G}({Guenther} 1994).

\revision{Six linear non-adiabatic equations are solved which take only
into account radiative losses and gains. The convective
flux is ``frozen" out of the pulsation equations
\markcite{1990ApJ...363..227P}(see {Pesnell} 1990, for a description on the
various ways in which this can be done)
thus the coupling of convection and the oscillations is not
accounted for. The calculated frequencies
are shown in \Fig\ref{fig:ech1}.}
Only one of the
eight frequencies in the high frequency regime matches
the ground based data points
reported by \markcite{2003AJ....126.1483K}{Kjeldsen} {et~al.} (2003), on average
the model yields folded frequencies about $3\,\mu$Hz larger.

Next we include the effects of turbulence in our
model according to \Sec\ref{sec:turb}. 
The free parameters mixing length and hydrogen abundance
are adjusted slightly
to ensure that the luminosity and effective temperature of the model
with turbulence matched the luminosity and effective temperature of the
model without turbulence.
The calibration procedure is stopped after luminosity
and effective temperature match with a relative difference
better than $5 \times 10^{-5}$.

\subsection{Radial modes}
Finally, the non-adiabatic $l=0,1,2\,$ $p$-mode frequency spectrum
is calculated with JIG and we plot the results in an
echelle diagram (\Fig\ref{fig:ech1}). We can see
in this figure that the model with turbulence still
matches the $l=0$ low frequency MOST data points as
required by our calibration technique while in addition
it reproduces six out of the eight ground based $l=0$
frequency data points by Kjeldsen, five within their $1\sigma$ uncertainty
and one within $2\sigma$. Also note that one frequency not matching
the data is still a match within $3\sigma$ and
that we are using the errors as quoted by
\markcite{2003AJ....126.1483K}{Kjeldsen} {et~al.} (2003).

The region of  $600-650\,\mu$Hz where the models coincide
with the two MOST modes that have been
independently confirmed by the ground based measured modes
of both \markcite{2003AJ....126.1483K}{Kjeldsen} {et~al.} (2003) and \markcite{2005A&A...434.1085C}{Carrier} {et~al.} (2005)
add credibility to our modeling.
As already noted in \markcite{2005ApJ...........G}{Guenther} {et~al.} (2005), MOST
had measured two modes below $200\,\mu$Hz that also fit into
the $l=0$ sequence of our models.

To provide a more quantitative measure we calculate the
$\chi^2$ numbers for the combined data sets (MOST plus
Kjeldsen et al.) of
all $l=0$ modes in the range $200-900\,\mu$Hz. For the
model without turbulence we get $\chi^2 = 18$ compared
to $\chi^2 = 2.5$ for the model that includes turbulence.

In \markcite{2005ApJ...........G}{Guenther} {et~al.} (2005) the best fit
to the combined MOST and Kjeldsen et al. modes, again only constrained
by the oscillation frequencies, yielded a $\chi^2 = 2.3$. But,
importantly, the model corresponding to this fit to the oscillation
data, did not fit \eboop's location in the HRD. By including turbulence
in our model, we fit the MOST oscillation data, the Kjeldsen et al.
oscillation data, and the observed position in the HRD
\revision{\markcite{2003A&A...404..341D,2005A&A...436..253T}({Di Mauro} {et~al.} 2003; {Th{\' e}venin} {et~al.} 2005)}
simultaneously.

Our model with turbulence
fits the Kjeldsen data
much better than the data from Carrier. The combined set
of MOST plus Carrier data gives a $\chi^2 = 18$. Since
the Carrier data appears consistently at lower folded frequencies,
the standard model without turbulence is very far off with
a $\chi^2 = 131$. Hence the model with turbulence is still
much closer to the Carrier data than a model without turbulence.

\begin{figure}[t]
\begin{center}
\includegraphics[height=0.46\textwidth,angle=-90]{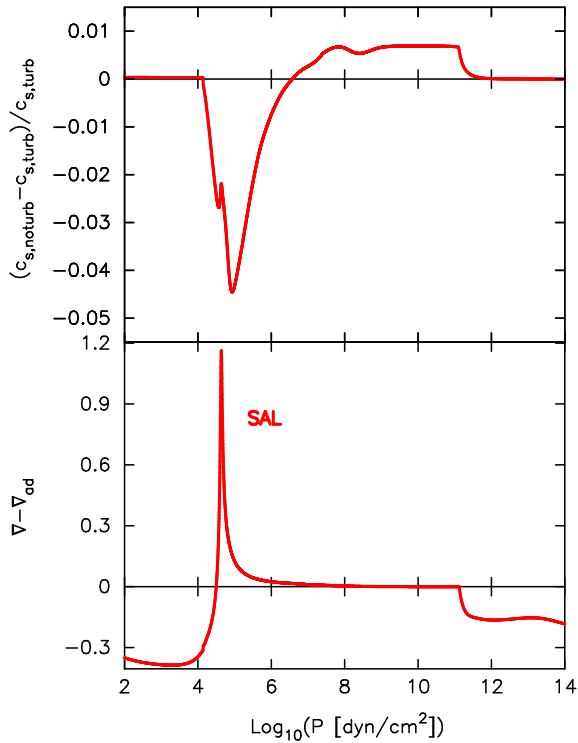}
\end{center}
\caption{Relative difference in sound speed of the
outer layers of \eboo between a model including and lacking
the effects of turbulence (top panel). The SAL of the model
including turbulence is shown for comparison (bottom panel).
\label{fig:sspeed}}
\end{figure}

The structural difference in sound speed between the model
including turbulence and without turbulence is shown in
\Fig\ref{fig:sspeed} (top panel). As expected, the largest
deviation of $\sim 4$\% is seen within the peak of the SAL (bottom panel).
Also, the deeper convective
layers are affected by about $1$\%. The structural differences
vanish at pressures greater than $10^{11}\,$dyn/cm$^2$, where the
layers are fully radiative.

\subsection{Nonradial modes}
Additional information is present in the $l=1,2$ modes.
From our models we expect some regular spacings for
the high frequency $p$-modes, but due to
\eboop's advanced evolutionary stage no regular spacing
should be seen for lower frequencies which are subject to
strong mode bumping. Unfortunately, MOST did not see many
high frequency modes and we must therefore mostly
rely on the ground based data set. 

We can see from
\Fig\ref{fig:ech1} that none of the five $l=2$ modes
from Kjeldsen are
matched within $2\sigma$ with the model lacking turbulence
whereas four matches are achieved with the model that includes
turbulence. The only mode that cannot be matched to
the Kjeldsen data is one with frequency greater than $950\,\mu$Hz.
Interestingly, we cannot reproduce
any of the three measurements above this
threshold (l=0,1,2) suggesting there is still room for
improvement in our models.
We identify one additional MOST mode
at $648\,\mu$Hz that fits smoothly into the high frequency
$l=2$ sequence. This mode is also better matched by the model including
turbulence.

The matches of the model with turbulence
to the $l=1$ ground based data are not as good as for $l=2$.
The model including turbulence
matches three out of eight measured frequencies in comparison
to only two by the model without turbulence. This ratio
is slightly enhanced when we include one additional
mode found with MOST at $828\,\mu$Hz. The more
successful fit of the $l=2$ modes can
be explained with the finding of \markcite{2005ApJ...........G}{Guenther} {et~al.} (2005)
who demonstrated that a slight perturbation in the model
mass of $0.005\,\Msun$ leads only to minimal changes in the
$l=0$ modes but to extreme changes in the $l=1,2$ modes
with particularly large impact on the $l=1$ modes. 

Of the many modes smaller than $500\,\mu$Hz that are
present in the MOST data we plot only those which lie near
modes predicted by the models. Although there are many
modes present in the MOST data very few modes are actually
matched by the models and no statistical advantage of the
models including turbulence can be inferred. Since
observations show no evidence for any
evenly spaced sequences in this frequency domain and models
predict mode bumping to occur as a result of
mixed $g$-modes for frequencies smaller than $350\,\mu$Hz and
mixed $p$-modes within $350-600\,\mu$Hz we are simply
not in the position yet to make use of this information 
in our models.

Of the nine nonradial modes measured by Carrier, none is
matched by the standard model frequencies and only one is
matched by a mode from the model that includes turbulence.
As noted before for the radial data,
the non radial $l=1,2$ appears at lower folded
frequencies. Again, models with turbulence fit better to the
Carrier data than models without turbulence but in either case there
remains a large discrepancy between these observations and the
models.

Finally, we list the frequencies from our theoretical models
in Table~\ref{tab:comp} together with the observed frequencies.
The observed modes are identified
with radial orders that most closely match our theoretical models.
Most of the measurements from the ground are
identified with a radial order $n$ one higher
than those in this work. The mixed mode character of all
modes is indicated by the number $n_g$.

\revision{
\subsection{Origin of improved p-mode frequencies}
The improvements of the fit between observed and model
high frequency p-modes arise from structural changes in the
superficial layers of the star, mainly in the SAL of the
convection zone. However, it is important to note that
the form of the structure change that reproduces the correct
shift of the high frequency p-modes is not unique.
For the Sun, \markcite{1996A&A...307..624M}{Monteiro}, {Christensen-Dalsgaard}, \&  {Thompson} (1996)
demonstrated that models with a steeper and narrower SAL compared
to standard MLT models can produce a frequency shift that brings
these models in accord with the observed p-mode frequencies.
In contrast to this, 3D simulations for the Sun
\markcite{1999ASPC..173...91N}({Nordlund} \& {Stein} 1999)
produce an SAL stratification that is very close to standard MLT.
The frequencies of high order p-modes are also predicted
smaller when calculated from the average structure of the 3D
simulations, hence, a better fit to observations is achieved.
\markcite{1999ASPC..173...91N}{Nordlund} \& {Stein} (1999) attribute this frequency shift
in their models
to the turbulent pressure support and, in addition, to 3D effects
arising from the net effect of the fluctuations of the opacity.

In the following, we explore how the inclusion of turbulence
as performed in this paper changes the surface layer structure of
\eboo and we try to identify the characteristic features that lead
to the correct p-mode frequency shifts. A similar analysis has been
given by \markcite{2002ApJ...567.1192L}{Li} {et~al.} (2002) for the Sun, here,
we extend this analysis to \eboo and
add some more information about the role played by the shape of the SAL.

\subsubsection{Turbulent Pressure}
As described in \Sec\ref{sec:turb}, we account for the effects of
turbulence on the stellar structure
by including the turbulent pressure and turbulent kinetic energy
taken from a 3D simulation for the Sun. Thus, we are able to explore
the relative importance of both effects to the correct shifting
of the high frequency p-modes. In order to do so, we calculate one
model for \eboo where we include the turbulent pressure alone and
completely omit the turbulent kinetic energy. This model is calibrated to
give the same luminosity and effective temperature and hence
radius as our previous models. In \Fig\ref{fig:ptonly} we compare
this model with turbulent pressure alone to the standard MLT model and the
model including the effects of both turbulent pressure and turbulent
kinetic energy.
\begin{figure}[t]
\begin{center}
\includegraphics[width=0.45\textwidth,angle=0]{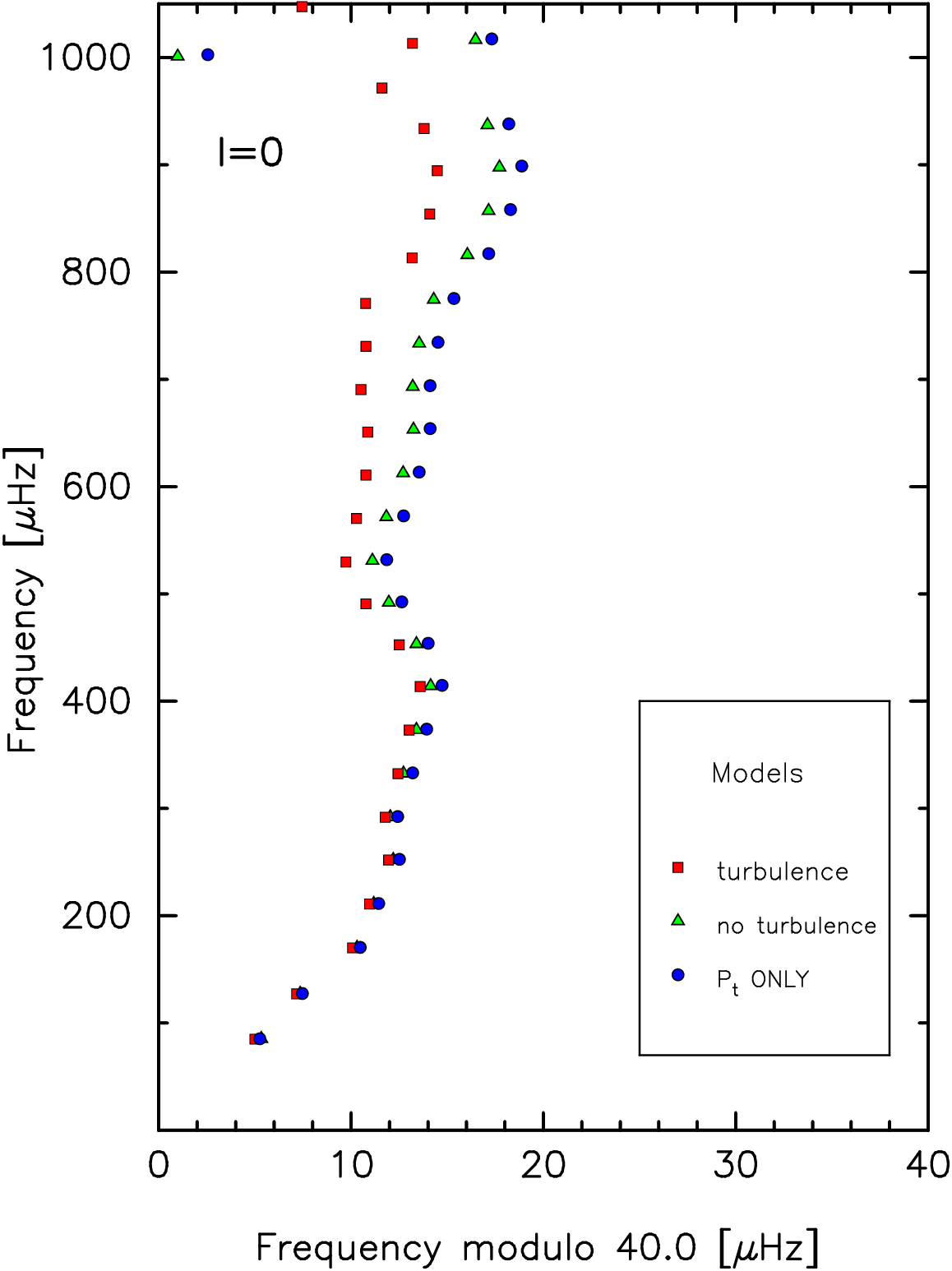}
\end{center}
\caption{Echelle diagram for a calibrated \eboo model where
only the effects of turbulent pressure are included (circles)
in comparison to the standard MLT model (triangles) and our
best model including the effects of turbulent pressure and
turbulent kinetic energy (squares).
\label{fig:ptonly}}
\end{figure}
As can be seen in \Fig\ref{fig:ptonly},
the model with turbulence alone shifts
the frequencies to higher folded frequencies, hence it increases the
discrepancy between the model and the observations. This finding is
supported by \markcite{2002ApJ...567.1192L}{Li} {et~al.} (2002, Fig.11), where the same
effect is seen in the case of the Sun. The same behavior has been found
for the Sun by \markcite{1992MNRAS.255..603B}{Balmforth} (1992, Table 1). 

\subsubsection{Turbulent Kinetic Energy}
It is obvious from our model with turbulence alone, which fails to
shift the p-mode frequencies into the right direction, that the main
ingredient for a better match with observations
is achieved by the effects of the turbulent kinetic energy. To show
this even more clearly we have calculated one additional model
in which we artificially
increased the turbulent kinetic energy by a factor of two.
Again, this model was properly calibrated.
As can bee seen in \Fig\ref{fig:doubledTKE}, increasing the
turbulent kinetic energy shifts the frequencies to lower folded
frequencies in the echelle diagram. 
\begin{figure}[t]
\begin{center}
\includegraphics[width=0.45\textwidth,angle=0]{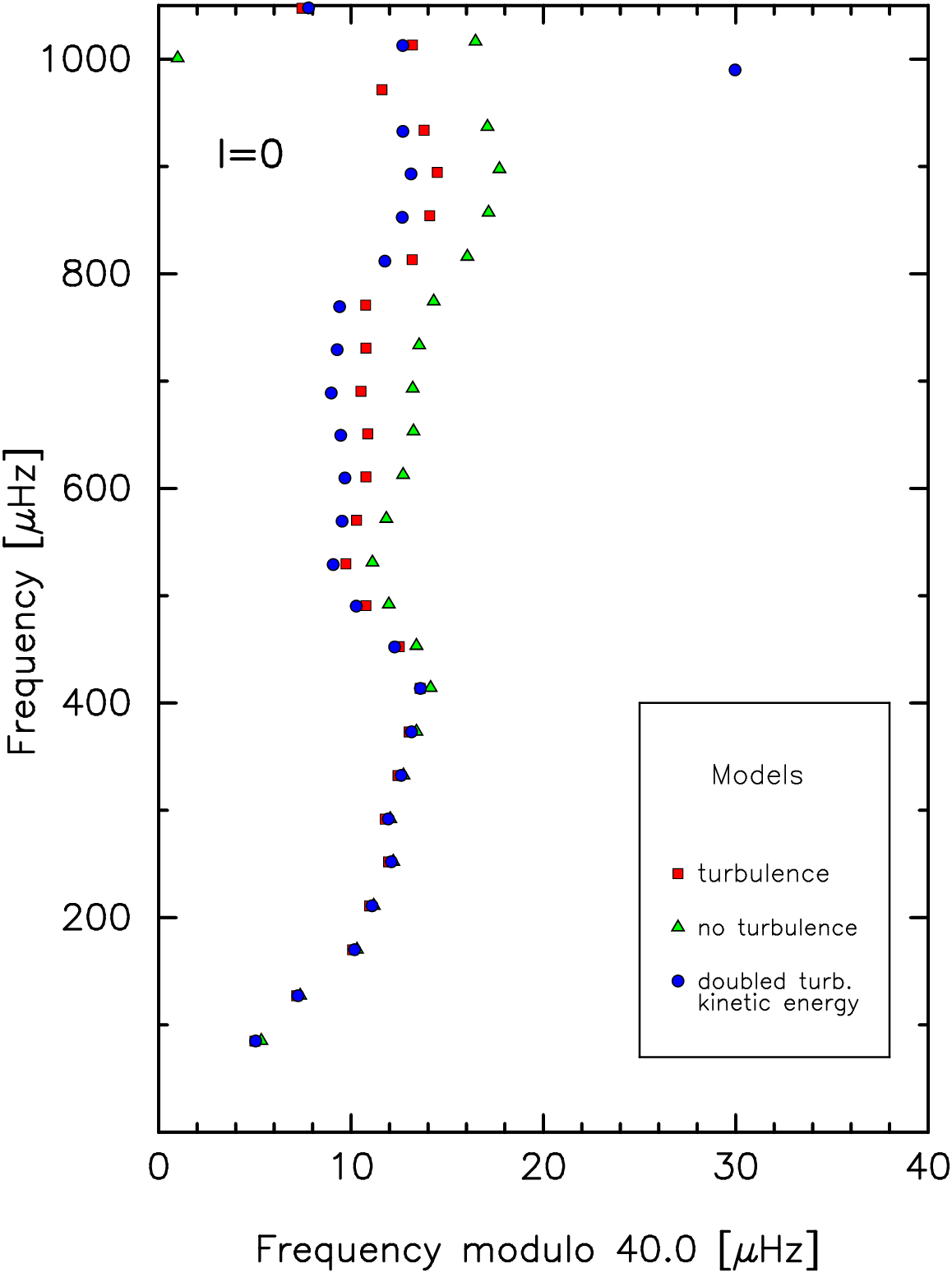}
\end{center}
\caption{Echelle diagram for a calibrated \eboo model where
the turbulent kinetic energy is doubled (circles)
in comparison to the standard MLT model (triangles) and our
best model (squares).
\label{fig:doubledTKE}}
\end{figure}

The plot also demonstrates that a quantitative match between observations
and models depends crucially on the exact magnitude of the turbulent
kinetic energy which we can only derive from a complete 3D simulation of
the outer layers of \eboo. Since we achieve a
good fit to the $p$-mode observations by 
applying to our 1D model the effects of turbulent
kinetic energy as derived from a 3D simulation for the Sun, it remains
to be shown that a 3D simulation of the outer layers of \eboo
yields comparable values for the turbulent kinetic energy.

\subsubsection{Shape of SAL}
If the effects of turbulent kinetic energy can shift the
model $p$-mode frequencies towards the observed frequencies, it is
illuminating to find out about the structural changes 
that are induced by the turbulent kinetic energy onto 
the outer layers of \eboop. These changes are best seen in the
SAL (\Fig\ref{fig:salvgl}).
\begin{figure}[t]
\begin{center}
\includegraphics[height=0.45\textwidth,angle=-90]{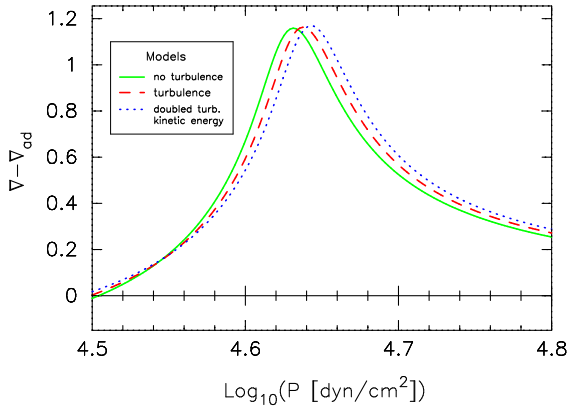}
\end{center}
\caption{The SAL for three different models: a standard MLT model
(solid line), our fiducial model with turbulent pressure and
turbulent kinetic energy included (dashed line) and a model
with artificially increased turbulent kinetic energy (dotted line).
\label{fig:salvgl}}
\end{figure}
The inclusion of turbulence has the effect of shifting the peak of
the SAL into deeper layers of the stellar envelope. This
is the main effect responsible for the frequency shift. Also,
the superadiabaticity is increased (increased peak height),
but this effect is small. It is worth noticing that
the shape of the SAL is preserved, with same half-maximum in all
models. In order to demonstrate this, we have shifted the SAL of
two models artificially to make
their peak location coincide with the standard MLT model
(\Fig\ref{fig:shape}).
\begin{figure}[t]
\begin{center}
\includegraphics[height=0.45\textwidth,angle=-90]{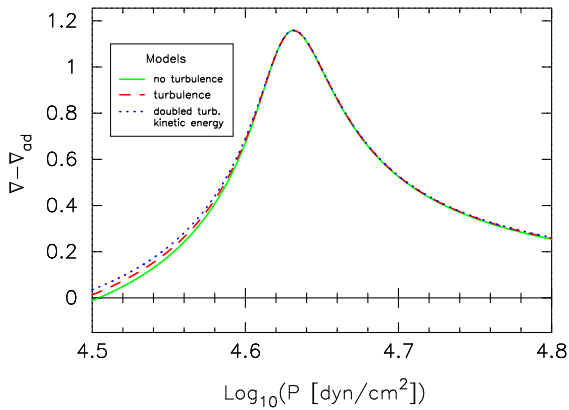}
\end{center}
\caption{Same plot as \Fig\ref{fig:salvgl} but the
models including turbulent kinetic energy have been shifted to
coincide with the peak of the standard MLT model to allow
for comparison of the SAL shape.
\label{fig:shape}}
\end{figure}
We conclude that the inclusion of turbulent kinetic energy shifts
the SAL peaks into deeper layers of the stellar envelope while
the shape of the SAL is preserved. The deeper SAL location
changes the run of sound speed in the outer layers leading to
high order $p$-mode frequency  shifts towards lower
folded frequencies as required by observational data for the Sun
and also for \eboop. For the latter, the exact same amount of
turbulent kinetic
energy as derived from a full 3D simulation for the Sun is able
to bring the models in accord with the high frequency
$p$-mode observations.
}

\section{\revision{Discussion and} Conclusions}
\label{sec:concl}
This paper demonstrates that the measured $p$-mode frequencies
of \eboo
from space (by MOST) and the ground (by Kjeldsen) can be \emph{jointly}
matched with our theoretical models
by including the effects of turbulence in the outer stellar layers.
We are able to
report on a better match between theory and observation
for a star other than the Sun
when the outer stellar layers are corrected by the effects
of turbulence. \revision{It is an important assumption of this paper,
that the effects of turbulence on the outer layer of \eboo
can be extracted from a 3D hydrodynamical
simulation of the surface layers of the Sun.}

\revision{
The high order $p$-mode frequency shift, that brings our model
in better agreement with the observations, is shown to be a direct
consequence of the inclusion of turbulent kinetic energy.
Although the turbulent kinetic energy must be present in 
the outer convection zone, it is usually disregarded in
traditional stellar modeling. The turbulent kinetic energy
in this study was taken from a 3D hydrodynamical convection
simulations of the Sun, therefore, only a 3D simulation
for \eboo can add the final
proof that the amount of turbulent kinetic energy we use in this
paper is correct.}

Specifically, we show that the inclusion of turbulence
\revision{into our 1D stellar evolution models,
as derived from the solar 3D simulation data}, 
can account for the difference of $1.5 - 4\mu\,$Hz in the
echelle diagram between the ground based $l=0$ data points and the models
without turbulence. The quantitative agreement for the model
including turbulence is excellent for the
combined data set (MOST plus Kjeldsen) of the $l=0$ $p$-mode frequencies
between $200-900\,\mu$Hz with a $\chi^2 = 2.5$. 

The better agreement between the observations and the models
that include turbulence, is strengthened by comparing the
observed ground based $l=1,2$ modes. The model with turbulence matches
four out of nine $l=1$ and five out of six $l=2$ modes.
If we combine all $22$ observed ground based modes with
$14$ selected MOST modes, a model with turbulence
reproduces $27$ modes versus $14$ modes for a model
without turbulence (within $2\sigma$ error bars).

\revision{
MOST sees more spectrum peaks than shown in our echelle diagrams.
The models predict, that we should be seeing mixed modes. However,
the stellar origin has to be established with more certainty. This
will be possible with the scheduled re-observation
of \eboo by MOST in 2005. If some of the recurring peaks can
be identified as nonradial modes, this will give us excellent
additional information to test our models.}

\revision{Because our theoretical models match the Kjeldsen data,
they are not a good match to the Carrier data, since both
data sets do not overlap for the majority of modes
which is true for both radial and nonradial modes.} Regardless
of this discrepancy, frequencies from models that include turbulence
come much closer to the Carrier measurements.

\revision{
Certainly the most important advancement for this study is
the inclusion of the turbulence data derived from a full
3D turbulence simulation for \eboop. We are currently undertaking
this task and will report on it in the
future.}
Also, the QDG search
-- already being very extensive -- must be expanded
in the searched parameter space of hydrogen content, metallicity
and convective core overshoot.

The refinements of stellar evolution theory with regard to
turbulent convection in the outer layers has previously been motivated by
attempts to gain a better match between theory and
observation for the frequency spectrum of the Sun. This
study \revision{indicates}
that these refinements to the theoretical
models are also important for interpreting the observational
data of \eboop. We believe that the need for extremely
precise theoretical models will continue to
grow as more and more observational measurements
become available within the young field of asteroseismology. 

\acknowledgments
We would like to thank Sarbani Basu 
for stimulating discussions during many stages of this work.
This research was supported by NASA grant NAG5-13299 (CWS and PD), and
in part by the NASA EOS/IDS Program (FJR). DBG acknowledges
support from an operating research grant from NSERC of Canada.

\bibliography{}

\clearpage
\begin{deluxetable}{llllllll}
\tabletypesize{\scriptsize}
\tablecaption{Model Frequencies and Observed Frequencies\label{tab:comp}}
\tablewidth{0pt}
\tablehead{
  \colhead{\phantom{$l=000$\ldots}} & \colhead{} & \colhead{} &
  \multicolumn{2}{c}{Model Frequencies}  &
  \multicolumn{3}{c}{Observations}  \\[1mm]
  \colhead{Order} & \colhead{$n_p$} & \colhead{$n_g$} &
  \colhead{With Turbulence} & \colhead{Without Turbulence} &
  \colhead{MOST 2005\tablenotemark{a,b}}   & \colhead{Kjeldsen 2003} &
  \colhead{Carrier 2005\tablenotemark{c}}
}
\startdata
 $l=0$\dotfill & $\phantom{0}1$ & $\phantom{0}0$   & $\phantom{0}127.17$ & $\phantom{0}127.35$ & $126.66$/$127.91$         & $$                                  & $$                              \\
               & $\phantom{0}2$ & $\phantom{0}0$   & $\phantom{0}170.08$ & $\phantom{0}170.31$ & $171.32$\tablenotemark{d} & $$                                  & $$                              \\
               & $\phantom{0}3$ & $\phantom{0}0$   & $\phantom{0}210.96$ & $\phantom{0}211.18$ & $210.56$\tablenotemark{d} & $$                                  & $$                              \\
               & $\phantom{0}4$ & $\phantom{0}0$   & $\phantom{0}251.95$ & $\phantom{0}252.19$ & $251.79$\tablenotemark{d} & $$                                  & $$                              \\
               & $\phantom{0}5$ & $\phantom{0}0$   & $\phantom{0}291.78$ & $\phantom{0}292.04$ & $292.25$\tablenotemark{d} & $$                                  & $$                              \\
               & $\phantom{0}6$ & $\phantom{0}0$   & $\phantom{0}332.45$ & $\phantom{0}332.73$ & $333.17$\tablenotemark{d} & $$                                  & $$                              \\
               & $\phantom{0}7$ & $\phantom{0}0$   & $\phantom{0}373.02$ & $\phantom{0}373.40$ & $373.20$\tablenotemark{d} & $$                                  & $$                              \\
               & $\phantom{0}8$ & $\phantom{0}0$   & $\phantom{0}413.60$ & $\phantom{0}414.14$ & $414.01$\tablenotemark{d} & $$                                  & $$                              \\
               & $\phantom{0}9$ & $\phantom{0}0$   & $\phantom{0}452.51$ & $\phantom{0}453.40$ & $453.13$\tablenotemark{d} & $$                                  & $$                              \\
               & $10$           & $\phantom{0}0$   & $\phantom{0}490.78$ & $\phantom{0}491.96$ & $492.92$\tablenotemark{d} & $$                                  & $$                              \\
               & $11$           & $\phantom{0}0$   & $\phantom{0}529.73$ & $\phantom{0}531.11$ & $$                        & $$                                  & $533.0$\tablenotemark{e}        \\
               & $12$           & $\phantom{0}0$   & $\phantom{0}570.29$ & $\phantom{0}571.84$ & $$                        & $$                                  & $$                              \\
               & $13$           & $\phantom{0}0$   & $\phantom{0}610.78$ & $\phantom{0}612.71$ & $610.55$                  & $\phantom{0}611.0\pm0.5$\tablenotemark{e} & $610.6$\tablenotemark{e}        \\
               & $14$           & $\phantom{0}0$   & $\phantom{0}650.87$ & $\phantom{0}653.25$ & $650.37$                  & $\phantom{0}651.2\pm0.6$\tablenotemark{e} & $$                              \\
               & $15$           & $\phantom{0}0$   & $\phantom{0}690.51$ & $\phantom{0}693.21$ & $$                        & $\phantom{0}690.8\pm0.6$\tablenotemark{e} & $691.3$\tablenotemark{e}        \\
               & $16$           & $\phantom{0}0$   & $\phantom{0}730.78$ & $\phantom{0}733.54$ & $$                        & $\phantom{0}732.6\pm0.4$\tablenotemark{e} & $729.5$\tablenotemark{e}        \\
               & $17$           & $\phantom{0}0$   & $\phantom{0}770.76$ & $\phantom{0}774.30$ & $$                        & $$                                  & $769.4$\tablenotemark{e}        \\
               & $18$           & $\phantom{0}0$   & $\phantom{0}813.18$ & $\phantom{0}816.05$ & $$                        & $\phantom{0}813.1\pm0.4$\tablenotemark{e} & $809.2$\tablenotemark{e}        \\
               & $19$           & $\phantom{0}0$   & $\phantom{0}854.09$ & $\phantom{0}857.15$ & $$                        & $\phantom{0}853.6\pm0.3$\tablenotemark{e} & $$                              \\
               & $20$           & $\phantom{0}0$   & $\phantom{0}894.48$ & $\phantom{0}897.71$ & $$                        & $\phantom{0}894.2\pm0.6$\tablenotemark{e} & $891.6$\tablenotemark{e}        \\
               & $21$           & $\phantom{0}0$   & $\phantom{0}933.80$ & $\phantom{0}937.09$ & $$                        & $$                                  & $$                              \\
               & $22$           & $\phantom{0}0$   & $\phantom{0}971.60$ & $1000.99$           & $$                        & $\phantom{0}974.5\pm0.7$\tablenotemark{e} & $971.9$\tablenotemark{e}        \\[5mm]
 $l=1$\dotfill & $\phantom{0}1$ & $20/22$          & $\phantom{0}116.00$ & $\phantom{0}104.19$ & 
                  & $$                                  & $$                              \\ 
               & $\phantom{0}1$ & $18$             & $$                  & $\phantom{0}125.92$ & 
                  & $$                                  & $$                              \\
               & $\phantom{0}2$ & $17$             & $\phantom{0}137.60$ & $$                  & 
                  & $$                                  & $$                              \\
               & $\phantom{0}2$ & $16$             & $\phantom{0}142.30$ & $\phantom{0}140.52$ & 
                        & $$                                  & $$                              \\ 
               & $\phantom{0}2$ & $12/10$          & $\phantom{0}177.76$ & $\phantom{0}177.22$ & 
                  & $$                                  & $$                              \\
               & $\phantom{0}3$ & $12$             & $$                  & $\phantom{0}185.19$ & 
                  & $$                                  & $$                              \\
               & $\phantom{0}3$ & $11$             & $$                  & $\phantom{0}196.86$ & 
                  & $$                                  & $$                              \\
               & $\phantom{0}3$ & $10$             & $\phantom{0}214.53$ & $$                  & 
                  & $$                                  & $$                              \\
               & $\phantom{0}3$ & $\phantom{0}9$   & $\phantom{0}221.96$ & $\phantom{0}221.40$ & 
                  & $$                                  & $$                              \\
               & $\phantom{0}4$ & $\phantom{0}6/8$ & $\phantom{0}261.27$ & $\phantom{0}259.47$ & 
                  & $$                                  & $$                              \\ 
               & $\phantom{0}5$ & $\phantom{0}7$   & $\phantom{0}298.48$ & $\phantom{0}294.38$ & 
                  & $$                                  & $$                              \\
               & $\phantom{0}5$ & $\phantom{0}6$   & $\phantom{0}304.40$ & $\phantom{0}303.45$ & 
                  & $$                                  & $$                              \\ 
               & $\phantom{0}6$ & $\phantom{0}4/6$ & $\phantom{0}339.66$ & $\phantom{0}336.48$ & 
                  & $$                                  & $$                              \\ 
               & $\phantom{0}6$ & $\phantom{0}5$   & $$                  & $\phantom{0}345.58$ & 
                  & $$                                  & $$                              \\
               & $\phantom{0}7$ & $\phantom{0}3$   & $$                  & $\phantom{0}379.58$ & 
                  & $$                                  & $$                              \\ 
               & $\phantom{0}8$ & $\phantom{0}2$   & $\phantom{0}380.22$ & $$                  & 
                  & $$                                  & $$                              \\ 
               & $\phantom{0}8$ & $\phantom{0}5$   & $\phantom{0}404.03$ & $$                  & 
                  & $$                                  & $$                              \\ 
               & $\phantom{0}8$ & $\phantom{0}4/2$ & $\phantom{0}422.43$ & $\phantom{0}420.77$ & 
                  & $$                                  & $$                              \\ 
               & $\phantom{0}9$ & $\phantom{0}2$   & $\phantom{0}448.69$ & $\phantom{0}447.84$ & 
                  & $$                                  & $$                              \\
               & $10$           & $\phantom{0}2$   & $\phantom{0}477.37$ & $\phantom{0}476.78$ & 
                  & $$                                  & $$                              \\
               & $10$           & $\phantom{0}3$   & $$                  & $\phantom{0}503.67$ & 
                  & $$                                  & $$                              \\
               & $11$           & $\phantom{0}4$   & $\phantom{0}507.43$ & $$                  & 
                  & $$                                  & $512.2$                         \\
               & $11$           & $\phantom{0}3$   & $\phantom{0}524.14$ & $\phantom{0}521.11$ & 
                  & $$                                  & $$                              \\
               & $12$           & $\phantom{0}1$   & $\phantom{0}553.78$ & $\phantom{0}554.48$ & 
                  & $$                                  & $550.3$                         \\
               & $13$           & $\phantom{0}1/3$ & $\phantom{0}592.10$ & $\phantom{0}593.18$ & 
                  & $$                                  & $589.9$                         \\
               & $14$           & $\phantom{0}3$   & $\phantom{0}630.64$ & $\phantom{0}631.91$ & 
                  & $\phantom{0}629.4\pm0.3$                  & $625.7$                         \\
               & $14$           & $\phantom{0}2$   & $\phantom{0}664.90$ & $\phantom{0}658.58$ & 
                  & $\phantom{0}670.1\pm0.5$\tablenotemark{e} & $665.4$/$669.9$\tablenotemark{e}\\
               & $15$           & $\phantom{0}2$   & $\phantom{0}678.77$ & $\phantom{0}676.75$ & 
                  & $$                                  & $$                              \\
               & $16$           & $\phantom{0}2$   & $\phantom{0}711.75$ & $\phantom{0}713.97$ & 
                  & $\phantom{0}711.8\pm0.4$                  & $$                              \\
               & $17$           & $\phantom{0}2$   & $\phantom{0}751.13$ & $\phantom{0}753.68$ & 
                  & $\phantom{0}749.3$/$753.4\pm0.5$          & $748.5$                         \\
               & $18$           & $\phantom{0}2$   & $$                  & $\phantom{0}795.64$ & 
                  & $\phantom{0}793.1\pm0.7$                  & $787.4$                         \\
               & $19$           & $\phantom{0}2$   & $\phantom{0}832.52$ & $\phantom{0}835.22$ & 
                  & $$                                  & $$                              \\
               & $20$           & $\phantom{0}2$   & $\phantom{0}872.54$ & $\phantom{0}875.18$ & 
                  & $$                                  & $$                              \\
               & $20$           & $\phantom{0}1$   & $\phantom{0}911.33$ & $\phantom{0}913.13$ & 
                  & $$                                  & $$                              \\
               & $21$           & $\phantom{0}1$   & $\phantom{0}940.11$ & $\phantom{0}927.53$ & 
                  & $$                                  & $$                              \\
               & $22$           & $\phantom{0}1$   & $\phantom{0}955.52$ & $\phantom{0}956.82$ & 
                  & $\phantom{0}955.6\pm0.8$                  & $947.6$                         \\
               & $23$           & $\phantom{0}1$   & $\phantom{0}994.58$ & $\phantom{0}997.34$ & 
                  & $$                                  & $$                              \\
               & $24$           & $\phantom{0}1$   & $1030.27$           & $1033.29$           & 
                  & $1034.3\pm0.7$                            & $$                              \\[5mm]
 $l=2$\dotfill & $\phantom{0}1$ & $30$             & $$                  & $\phantom{0}129.65$ & 
                  & $$                                  & $$                              \\
               & $\phantom{0}2$ & $25$             & $$                  & $\phantom{0}157.49$ & 
                  & $$                                  & $$                              \\
               & $\phantom{0}2$ & $23$             & $$                  & $\phantom{0}219.36$ & 
                  & $$                                  & $$                              \\
               & $\phantom{0}3$ & $19$             & $$                  & $\phantom{0}200.90$ & 
                  & $$                                  & $$                              \\
               & $\phantom{0}3$ & $18$             & $\phantom{0}208.41$ & $\phantom{0}206.27$ & 
                  & $$                                  & $$                              \\
               & $\phantom{0}3$ & $16$             & $$                  & $\phantom{0}224.96$ & 
                  & $$                                  & $$                              \\
               & $\phantom{0}4$ & $16$             & $\phantom{0}237.23$ & $$                  & 
                  & $$                                  & $$                              \\
               & $\phantom{0}4$ & $15/14$          & $\phantom{0}246.81$ & $\phantom{0}250.28$ & 
                  & $$                                  & $$                              \\
               & $\phantom{0}5$ & $13$             & $\phantom{0}285.04$ & $\phantom{0}283.02$ & 
                  & $$                                  & $$                              \\
               & $\phantom{0}5$ & $12$             & $\phantom{0}290.29$ & $\phantom{0}286.99$ & 
                  & $$                                  & $$                              \\
               & $\phantom{0}5$ & $11$             & $\phantom{0}310.59$ & $$                  & 
                  & $$                                  & $$                              \\
               & $\phantom{0}6$ & $11$             & $$                  & $\phantom{0}322.63$ & 
                  & $$                                  & $$                              \\
               & $\phantom{0}6$ & $10$             & $\phantom{0}331.04$ & $\phantom{0}328.29$ & 
                  & $$                                  & $$                              \\
               & $\phantom{0}6$ & $\phantom{0}9$   & $\phantom{0}354.29$ & $$                  & 
                  & $$                                  & $$                              \\
               & $\phantom{0}7$ & $\phantom{0}9$   & $\phantom{0}368.25$ & $\phantom{0}368.51$ & 
                  & $$                                  & $$                              \\
               & $\phantom{0}8$ & $\phantom{0}8$   & $\phantom{0}409.19$ & $\phantom{0}409.66$ & 
                  & $$                                  & $$                              \\
               & $\phantom{0}9$ & $\phantom{0}7$   & $\phantom{0}448.62$ & $\phantom{0}449.46$ & 
                  & $$                                  & $$                              \\
               & $\phantom{0}9$ & $\phantom{0}6$   & $$                  & $\phantom{0}477.88$ & 
                  & $$                                  & $$                              \\
               & $10$           & $\phantom{0}7$   & $\phantom{0}486.02$ & $$                  & 
                  & $$                                  & $$                              \\
               & $10$           & $\phantom{0}6$   & $\phantom{0}487.51$ & $\phantom{0}488.28$ & 
                  & $$                                  & $$                              \\
               & $11$           & $\phantom{0}6$   & $\phantom{0}526.09$ & $\phantom{0}527.45$ & 
                  & $$                                  & $$                              \\
               & $11$           & $\phantom{0}5$   & $\phantom{0}554.59$ & $$                  & 
                  & $$                                  & $$                              \\
               & $13$           & $\phantom{0}5$   & $\phantom{0}607.32$ & $\phantom{0}609.20$ & 
                  & $\phantom{0}608.1\pm0.4$                  & $$                              \\
               & $14$           & $\phantom{0}4$   & $\phantom{0}647.83$ & $\phantom{0}649.90$ & 
                  & $$                                  & $$                              \\
               & $15$           & $\phantom{0}4$   & $\phantom{0}687.22$ & $\phantom{0}689.95$ & 
                  & $$                                  & $$                              \\
               & $16$           & $\phantom{0}4$   & $\phantom{0}727.43$ & $\phantom{0}730.17$ & 
                  & $\phantom{0}728.4\pm0.6$                  & $724.5$                         \\
               & $17$           & $\phantom{0}3$   & $\phantom{0}775.69$ & $\phantom{0}770.39$ & 
                  & $$                                  & $765.6$                         \\
               & $18$           & $\phantom{0}3$   & $\phantom{0}810.81$ & $\phantom{0}798.35$ & 
                  & $\phantom{0}810.5\pm0.4$                  & $805.1$                         \\
               & $19$           & $\phantom{0}3$   & $\phantom{0}850.78$ & $\phantom{0}853.81$ & 
                  & $\phantom{0}849.9\pm0.7$                  & $846.1$                         \\
               & $20$           & $\phantom{0}3$   & $\phantom{0}891.24$ & $\phantom{0}894.42$ & 
                  & $$                                  & $888.7$                         \\
               & $21$           & $\phantom{0}3$   & $\phantom{0}930.84$ & $\phantom{0}933.81$ & 
                  & $$                                  & $$                              \\
               & $22$           & $\phantom{0}2$   & $\phantom{0}962.48$ & $\phantom{0}965.38$ & 
                  & $\phantom{0}971.7\pm0.6$                  & $$                              \\
\enddata

\tablenotetext{a}{We do not list the $l=1,2$ MOST modes until they are confirmed.}
\tablenotetext{b}{The observational uncertainty is $\pm0.40\mu$Hz (quoted from original work).}
\tablenotetext{c}{The observational uncertainty is $\pm0.44\mu$Hz (quoted from original work).}
\tablenotetext{d}{Modes used in QDG search.}
\tablenotetext{e}{Identified as radial order $n+1$ in original work.}


\end{deluxetable}

\end{document}